\documentclass[sigconf,nonacm]{acmart}

\usepackage{graphicx}
\usepackage{subcaption}
\usepackage{tikz}
\usetikzlibrary{positioning,shapes.geometric,shapes.symbols,shadows,arrows.meta,fit,calc}

\AtBeginDocument{}

\setcopyright{none}
\acmConference{}{}{}
\acmBooktitle{}
\acmDOI{}
\acmISBN{}
\acmYear{}
\copyrightyear{}
\date{}

\hypersetup{hidelinks}

\fancypagestyle{preprint}{%
  \fancyhf{}%
  \fancyfoot[C]{\normalfont\footnotesize\thepage}%
}

\begin{document}

\title[AdaFrame: Adaptive Video Ad Analysis]{%
  AdaFrame: Hierarchical Multimodal Deduplication%
  \texorpdfstring{\\}{ }with Adaptive Information Budgeting for%
  \texorpdfstring{\\}{ }Cost-Efficient Video Advertisement Analysis}


\author{Abdul Basit Tonmoy}
\affiliation{%
  \institution{Wabash College}
  \city{Crawfordsville}
  \state{Indiana}
  \country{USA}
}
\email{atonmoy27@wabash.edu} 

\author{Arman Luthra}
\affiliation{%
  \institution{Wabash College}
  \city{Crawfordsville}
  \state{Indiana}
  \country{USA}
}
\email{aluthra26@wabash.edu} 

\author{Guang Lin}
\authornote{Guang Lin and Qixin Deng are co-corresponding authors.}
\affiliation{%
  \institution{Purdue University}
  \city{West Lafayette}
  \state{Indiana}
  \country{USA}
}
\email{guanglin@purdue.edu} 

\author{Qixin Deng}
\authornotemark[1]
\affiliation{%
  \institution{Wabash College}
  \city{Crawfordsville}
  \state{Indiana}
  \country{USA}
}
\email{dengq@wabash.edu} 

\renewcommand{\shortauthors}{Tonmoy et al.}

\begin{abstract}
Uniform frame sampling for VLM-based video analysis floods multimodal backbones with redundant content, while existing keyframe methods optimize for human consumption rather than downstream extraction accuracy. We propose \textbf{AdaFrame}, a content-adaptive preprocessing cascade that reduces visual redundancy without sacrificing extraction fidelity. Three progressively finer filters---perceptual hashing, learned perceptual similarity (LPIPS), and CLIP-based semantic clustering---eliminate redundancy cheaply before committing to costly cross-modal inference. Per-video frame budgets are set by \emph{Intrinsic Semantic Dimensionality} (ISD), which estimates each video's semantic complexity via SVD of its frame embeddings. On 500 video advertisements from the Pitt Ads Dataset, AdaFrame reduces frame count by 48\% and VLM inference cost by 75\% vs.\ uniform sampling, while achieving 86.0\% topic accuracy (+1.6pp) and 79.4\% sentiment accuracy (+7.5pp). Code and evaluation framework will be released upon publication.
\end{abstract}



\maketitle
\pagestyle{preprint}
\thispagestyle{preprint}

\section{Introduction}
\label{sec:introduction}

The proliferation of video advertising has created an unprecedented demand for automated content analysis. Brands need to extract structured metadata---product categories, brand mentions, emotional tone, call-to-action elements---from millions of hours of ad content for competitive intelligence, compliance monitoring, and creative optimization. Vision-Language Models (VLMs) such as GPT-4V, Gemini, and Claude have demonstrated remarkable capability in this task, achieving near-human accuracy in identifying subtle brand placements and contextual nuances~\cite{liu2023visual, team2023gemini}.

However, VLM inference remains prohibitively expensive at scale. Processing a 30-second advertisement at 30 FPS requires analyzing 900 frames; even with aggressive subsampling, costs accumulate rapidly across large corpora. Current approaches employ uniform frame sampling (e.g., 1 FPS or every $k$-th frame), which suffers from two fundamental limitations: (1) \textit{redundancy}---static scenes or slow pans generate near-identical frames that waste inference budget, and (2) \textit{blindness}---fixed intervals may miss rapid cuts containing critical brand reveals or product demonstrations.

Recent work on keyframe extraction~\cite{potapov2014category, singh2022temporal} and video summarization~\cite{zhou2018deep, zhao2021video} addresses redundancy but focuses on human consumption rather than machine analysis. These methods optimize for visual appeal or narrative coherence, not information preservation for downstream VLM tasks. Moreover, they lack principled methods for determining \textit{how many} frames suffice for a given video's semantic complexity.

We propose a \textbf{three-tier hierarchical deduplication cascade} that progressively filters redundant frames with increasing computational cost and finer semantic granularity:
\begin{enumerate}
    \item \textbf{Hash Voting}: Ultra-fast perceptual hashes (pHash, dHash, wHash) eliminate exact/near-duplicate frames with $O(1)$ comparison cost.
    \item \textbf{Perceptual Similarity}: LPIPS metrics capture perceptual similarity missed by hashes, handling lighting changes and minor transformations.
    \item \textbf{Semantic Clustering}: CLIP embeddings group semantically equivalent frames, with temporal-aware Non-Maximum Suppression ensuring coverage across the video timeline.
\end{enumerate}

Frame budgets are determined adaptively via \textbf{Intrinsic Semantic Dimensionality (ISD)}, computed through SVD of frame embeddings, combined with a \textbf{Semantic Energy} measure that captures both temporal change rate and spatial information density. This formulation provides an interpretable budget allocation algorithm rather than a fixed heuristic. Our contributions are:
\begin{itemize}
    \item \textbf{Hierarchical Deduplication Cascade}: A three-tier architecture (hash voting $\rightarrow$ LPIPS $\rightarrow$ CLIP clustering) that reduces VLM input frames by 48\% compared to uniform sampling, with temporal-aware NMS ensuring semantic coverage across the video timeline.
    \item \textbf{Adaptive Budget Estimation}: An SVD-based ISD measure, combined with Semantic Energy scaling, that determines per-video frame budgets based on semantic complexity. Validated on 500 videos ($r = 0.859$ between ISD and cut frequency, $p < 0.001$).
    \item \textbf{Provider Generality}: Validation across two VLM backends (Gemini~2.5~Flash and Claude~3~Haiku) confirming that AdaFrame's frame selection transfers across providers with ${<}1$pp variation in all accuracy dimensions, demonstrating the pipeline's model-agnostic design.
\end{itemize}
\section{Related Work}
\label{sec:related_work}
\subsection{Video Frame Selection and Summarization}
Traditional keyframe extraction methods fall into three categories: shot-based~\cite{potapov2014category}, clustering-based~\cite{hanjalic2004unified}, and optimization-based~\cite{das2011video}. Shot detection identifies scene boundaries via color histogram differences~\cite{cernekova2006entropy} or motion analysis~\cite{trivedi2007multi}, then selects representative frames per shot. While efficient, these methods ignore semantic content and struggle with gradual transitions common in advertisements.
Deep learning approaches have improved summarization quality. Zhou et al.~\cite{zhou2018deep} introduced DSNet for diverse video summarization using reinforcement learning. Zhao et al.~\cite{zhao2021video} proposed VSNet with attention mechanisms for highlight detection. However, these methods optimize for human viewing preferences (coverage, diversity, representativeness) rather than machine analysis fidelity. Recent work on task-aware summarization~\cite{rochan2023task} shows promise but lacks theoretical bounds on information preservation.
\subsection{Perceptual Hashing and Similarity Metrics}
Perceptual hashing algorithms (pHash~\cite{zauner2010framework}, dHash~\cite{krawczyk2015phash}, wHash~\cite{wang2019robust}) provide efficient fingerprinting for image deduplication. Learned perceptual metrics (LPIPS~\cite{zhang2018unreasonable}), trained on human judgment data, correlate better with perceived similarity than pixel-wise metrics like SSIM~\cite{wang2004image}. Our cascade uses hashes for speed and LPIPS for perceptual precision.
\subsection{Vision-Language Models for Video Analysis}
VLMs have revolutionized multimodal understanding. GPT-4V~\cite{openai2023gpt4v} demonstrates emergent capabilities in scene understanding. Gemini~\cite{team2023gemini} supports native video input with temporal reasoning. Recent work addresses VLM efficiency through token pruning~\cite{chen2024tokenpacker} and cache reuse~\cite{kwon2023efficient}. However, these methods operate at the model architecture level, orthogonal to our input-level optimization. Our approach complements these by reducing the number of forward passes required.
\subsection{SVD-Based Dimensionality Estimation}
The concept of Intrinsic Dimension via SVD has applications in neural network compression~\cite{denton2014exploiting} and manifold learning~\cite{fefferman2016testing}. We adapt intrinsic dimension to video semantics, defining ISD as the effective rank of the frame embedding matrix, and combine it with a composite budget scaling that captures temporal dynamics.
\section{System Overview}
\label{sec:system_overview}

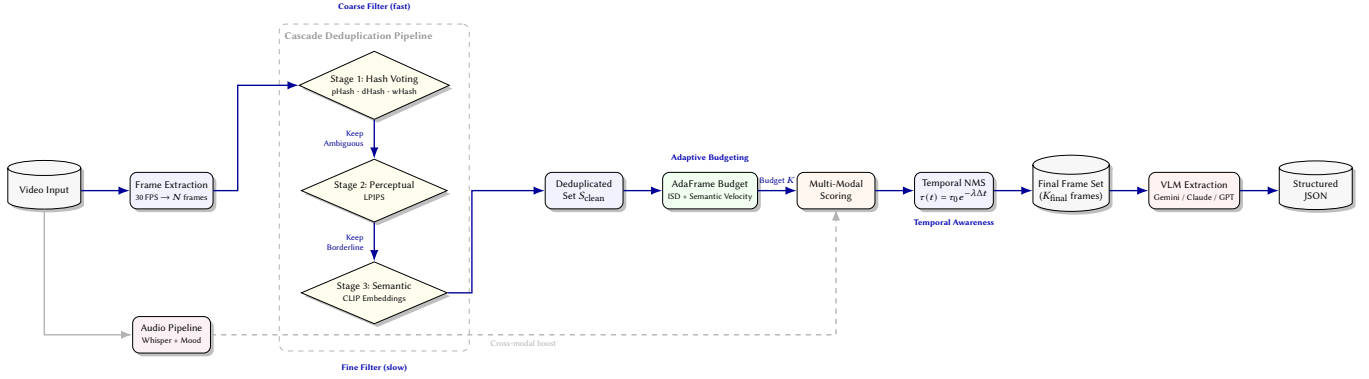
\begin{figure*}[t]
\centering

\resizebox{\textwidth}{!}{%
\begin{tikzpicture}[
    node distance=0.5cm and 0.8cm,
    >=Latex,
    font=\sffamily\scriptsize,
    box/.style={
        draw, rounded corners, align=center,
        minimum height=0.75cm, minimum width=1.6cm,
        fill=white, drop shadow
    },
    process/.style={box, fill=blue!5},
    decision/.style={
        draw, diamond, aspect=2.2,
        align=center, inner sep=1.5pt,
        minimum width=3.0cm, minimum height=0.85cm,
        fill=yellow!8, drop shadow,
        font=\sffamily\scriptsize
    },
    data/.style={
        draw, cylinder, shape border rotate=90, aspect=0.25,
        align=center, minimum height=1.0cm, minimum width=1.5cm,
        fill=gray!6, drop shadow,
        font=\sffamily\scriptsize
    },
    group/.style={
        draw=gray!60, dashed, rounded corners=4pt, inner sep=0.40cm,
        label={[anchor=north west, font=\sffamily\bfseries\scriptsize,
                text=gray!80, yshift=-0.05cm]north west:#1}
    },
    arrow/.style={->, thick, gray!60},
    highlight/.style={->, thick, blue!60!black},
    annotlabel/.style={font=\sffamily\tiny\bfseries, blue!70!black}
]

\node[data] (input) {Video Input};

\node[process, right=1.0cm of input] (extract)
    {Frame Extraction\\\tiny 30\,FPS $\to$ $N$ frames};
\draw[highlight] (input) -- (extract);

\node[decision, right=1.8cm of extract] (stage2)
    {Stage 2: Perceptual\\\tiny LPIPS};
\node[decision, above=0.8cm of stage2] (stage1)
    {Stage 1: Hash Voting\\\tiny pHash $\cdot$ dHash $\cdot$ wHash};
\node[decision, below=0.8cm of stage2] (stage3)
    {Stage 3: Semantic\\\tiny CLIP Embeddings};

\node[group={Cascade Deduplication Pipeline},
      fit={(stage1) (stage2) (stage3)},
      inner ysep=0.55cm] (cascade_group) {};

\node[annotlabel, above=0.15cm of cascade_group.north, anchor=south] {Coarse Filter (fast)};
\node[annotlabel, below=0.15cm of cascade_group.south, anchor=north] {Fine Filter (slow)};

\draw[highlight] (extract.east) -- ++(0.5,0) |- (stage1.west);

\draw[highlight] (stage1.south) --
    node[left, font=\sffamily\tiny, align=right, xshift=-0.1cm] {Keep\\Ambiguous} (stage2.north);
\draw[highlight] (stage2.south) --
    node[left, font=\sffamily\tiny, align=right, xshift=-0.1cm] {Keep\\Borderline} (stage3.north);

\node[process, right=2.0cm of stage2] (deduped)
    {Deduplicated\\Set $S_\mathrm{clean}$};

\draw[highlight] (stage3.east) -- ++(0.5,0) |- (deduped.west);

\node[process, right=0.8cm of deduped, fill=green!6] (hib)
    {AdaFrame Budget\\\tiny ISD + Semantic Velocity};
\draw[highlight] (deduped) -- (hib);
\node[annotlabel, above=0.1cm of hib] {Adaptive Budgeting};

\node[process, right=0.8cm of hib, fill=orange!6] (scoring)
    {Multi-Modal\\Scoring};
\draw[highlight] (hib) -- node[above, font=\sffamily\tiny] {Budget $K$} (scoring);

\node[process, right=0.8cm of scoring] (nms)
    {Temporal NMS\\\tiny $\tau(t)=\tau_0 e^{-\lambda\Delta t}$};
\draw[highlight] (scoring) -- (nms);
\node[annotlabel, below=0.1cm of nms] {Temporal Awareness};

\node[data, right=0.8cm of nms] (output)
    {Final Frame Set\\($K_\mathrm{final}$ frames)};
\draw[highlight] (nms) -- (output);

\node[process, right=0.8cm of output, fill=red!5] (vlm)
    {VLM Extraction\\\tiny Gemini / Claude / GPT};
\draw[highlight] (output) -- (vlm);

\node[data, right=0.8cm of vlm] (json) {Structured\\JSON};
\draw[highlight] (vlm) -- (json);

\node[process, below=2.2cm of extract, fill=purple!6] (audio)
    {Audio Pipeline\\\tiny Whisper + Mood};
\draw[arrow] (input.south) |- (audio.west);
\draw[arrow, dashed] (audio.east) -|
    node[pos=0.25, below, font=\sffamily\tiny] {Cross-modal boost}
    (scoring.south);

\end{tikzpicture}%
}
\caption{System architecture. Videos flow through parallel audio-visual pipelines. The visual pipeline applies three-tier deduplication (hash voting $\rightarrow$ LPIPS $\rightarrow$ CLIP clustering) with adaptive budget allocation via ISD and Semantic Energy. Selected frames undergo OCR pre-processing before single-pass VLM extraction. Audio pipeline transcribes speech and classifies mood. Final outputs merge both modalities.}
\label{fig:architecture}
\end{figure*}

AdaFrame processes video advertisements through a seven-stage pipeline organized into two parallel tracks that merge before final extraction (Figure~\ref{fig:architecture}).

\subsection{Pipeline Architecture}

\textbf{Stage 1: Video Ingestion.} The input video is loaded and metadata extracted (duration, resolution, FPS, total frame count). Audio is simultaneously extracted to a separate file for parallel processing.

\textbf{Stages 2--4: Visual Pipeline} (runs in parallel with Stage 5).

\begin{itemize}
    \item \textit{Stage 2: Scene Detection.} Content-aware scene boundaries are detected via adaptive thresholding on inter-frame differences (PySceneDetect~\cite{castellano2022pyscenedetect} with content detector, threshold $= 27.0$). A fallback mechanism activates when fewer than 2 scenes are detected: the threshold is halved, and if still insufficient, the video is split into artificial chunks of 10\,s each. This ensures every video produces a usable scene segmentation regardless of editing style.

    \item \textit{Stage 3: Candidate Frame Extraction.} Within each scene, candidate frames are extracted at 50\,ms intervals using histogram-based change detection (chi-square distance, threshold $= 0.15$). A minimum temporal gap of 100\,ms between candidates prevents over-sampling during rapid motion. This stage reduces a typical 30\,s video from $\sim$900 native frames to $\sim$150 candidates.

    \item \textit{Stage 4: Hierarchical Deduplication.} The three-tier cascade (Section~\ref{sec:cascade}) progressively filters candidates: hash voting eliminates near-duplicates (37.5\% reduction), LPIPS removes perceptually similar frames (19.5\% further reduction), and CLIP clustering groups semantically equivalent content (53.3\% further reduction). Frames surviving all three tiers advance to selection.
\end{itemize}

\textbf{Stage 5: Audio Pipeline} (runs in parallel with Stages 2--4).
Audio is processed through four sub-components:
\begin{itemize}
    \item \textit{Speech detection}: Voice Activity Detection (VAD) identifies speech segments; if no speech is detected, transcription is skipped entirely to save compute.
    \item \textit{Transcription}: Whisper-large-v3~\cite{radford2022whisper} produces timestamped speech segments with word-level alignment.
    \item \textit{Key phrase extraction}: Promotional keywords (``sale'', ``free'', ``limited'', ``call now'') are detected in transcriptions with timestamps, enabling cross-modal alignment.
    \item \textit{Mood classification}: Audio features (RMS energy, spectral centroid, tempo) are used to classify mood as energetic, upbeat, calm, dramatic, or melancholic.
\end{itemize}

The audio context is formatted as a structured prompt supplement: transcription segments with timestamps, detected key phrases, and mood classification are appended to the VLM extraction prompt, enabling the model to cross-reference visual and verbal content.

\textbf{Stage 6: Adaptive Selection.} The visual and audio pipelines merge. An importance scorer assigns each surviving candidate a score based on: video position (opening/closing segments boosted 1.4--1.5$\times$), scene boundary proximity (1.2--1.4$\times$), audio event alignment (1.3$\times$ near energy peaks, 1.5$\times$ near key phrases), and visual features (1.3$\times$ for text overlays, 1.2$\times$ for faces). The adaptive budget (Section~\ref{sec:budget}) determines how many frames to keep, and the top-scoring candidates are selected with temporal-aware NMS (Section~\ref{sec:tanms}) ensuring temporal diversity.

\textbf{Stage 7: VLM Extraction.} Selected frames are sent to a VLM extraction backend (Gemini~2.5~Flash~\cite{team2023gemini} in primary experiments) with a structured extraction prompt requesting JSON output in a single pass. The prompt covers eleven field groups---brand identity, product, promotion, CTA, message, visuals, content rating, topic, sentiment, engagement metrics, and persuasion techniques---with topic and sentiment constrained to the closed label sets of the Pitt Ads Dataset (38 and 30 classes respectively), ensuring predicted IDs are directly comparable to human-annotated ground truth. The prompt also includes temporal context (frame timestamps, inter-frame deltas, position labels) and the audio context from Stage~5. A representative extraction output is shown in Table~\ref{tab:example}.

\subsection{Data Processing}
\label{sec:data_processing}

\textbf{Frame Representation.} All frames are resized to a maximum resolution of 720p (preserving aspect ratio) before processing. CLIP ViT-B/32 embeddings (512-dimensional) are computed in batches of 32 on GPU for the deduplication and clustering stages.

\textbf{Parallelization.} Stages 2--4 (visual) and Stage 5 (audio) execute concurrently via a thread pool. For batch processing of multiple videos, a process-level parallel pipeline distributes videos across workers, with each worker running the full seven-stage pipeline independently. Configuration, model weights (CLIP, Whisper), and VLM client connections are initialized once per worker.

\textbf{Output Format.} Each processed video produces a structured JSON result containing: video metadata (duration, FPS, resolution), scene boundaries, selected frame timestamps with importance scores, per-tier deduplication statistics (frame counts after hash/LPIPS/CLIP), audio context summary, and the full VLM extraction output. Batch results are saved incrementally with resume support---interrupted runs can be continued without reprocessing completed videos.

\textbf{Worked Example.} Table~\ref{tab:example} traces a representative 30-second gaming advertisement through the full pipeline, illustrating the progressive frame reduction at each stage and the final structured extraction.

\begin{table}[t]
\centering
\caption{End-to-end processing example for a 30-second gaming advertisement (``Sword of Chaos'', 30\,FPS, 981 total frames).}
\label{tab:example}
\resizebox{\columnwidth}{!}{%
\begin{tabular}{@{}lrl@{}}
\toprule
\textbf{Stage} & \textbf{Frames} & \textbf{Notes} \\
\midrule
\multicolumn{3}{@{}l}{\textit{Input}} \\
\quad Native video       & 981  & 30\,FPS $\times$ 32.7\,s \\
\midrule
\multicolumn{3}{@{}l}{\textit{Visual Pipeline}} \\
\quad Scene detection    & ---  & 8 scenes detected \\
\quad Candidate extraction & 448 & Histogram change detection at 50\,ms \\
\quad After pHash        & 280  & 37.5\% removed (near-duplicates) \\
\quad After LPIPS        & 225  & 19.6\% removed (perceptual similarity) \\
\quad After CLIP         & 21   & 90.7\% removed (semantic clustering) \\
\midrule
\multicolumn{3}{@{}l}{\textit{Audio Pipeline (parallel)}} \\
\quad Speech segments    & ---  & 3 segments, 12.4\,s total speech \\
\quad Key phrases        & ---  & None detected \\
\quad Mood               & ---  & Energetic \\
\midrule
\multicolumn{3}{@{}l}{\textit{Selection}} \\
\quad ISD ($k^*$, $\tau{=}0.90$) & --- & $k^* = 43$, budget $= 24$ \\
\quad Final selected     & 24   & Top-scored with TA-NMS \\
\midrule
\multicolumn{3}{@{}l}{\textit{Compression}} \\
\quad Ratio              & \multicolumn{2}{l}{$981 / 24 = 40.9\times$} \\
\quad VLM cost           & \multicolumn{2}{l}{\$0.07 (vs.\ \$0.55 for Uniform-1FPS)} \\
\midrule
\multicolumn{3}{@{}l}{\textit{VLM Extraction Output}} \\
\quad Ad type            & ---  & Entertainment \\
\quad Brand              & ---  & ``Sword of Chaos'', logo at 28.2\,s (high contrast) \\
\quad Topic              & ---  & ID 21: Games and toys (high confidence) \\
\quad Sentiment          & ---  & Active (ID 1); secondary: Confident, Fashionable, Youthful \\
\quad Engagement         & ---  & Exciting: 0.9, Funny: 0.0, Effectiveness: 4/5 \\
\quad Target audience    & ---  & 18--25; anime, fantasy games, action RPGs, mobile gaming \\
\quad Promotion / CTA    & ---  & None detected \\
\quad NSFW               & ---  & True \\
\quad Persuasion         & ---  & Visual appeal, emotion \\
\quad Confidence         & ---  & 0.77 \\
\bottomrule
\end{tabular}%
}
\end{table}

The extraction output confirms correct topic classification (ID 21, matching ground truth) and captures the ad's dominant tone---high excitement, no humor, strong visual appeal targeting a young gaming audience. The absence of any promotional offer or CTA is consistent with an entertainment-style game trailer focused on brand awareness rather than direct response. The audio pipeline contributes mood classification (energetic) but detects no promotional keywords, corroborating the extraction result.
\section{Methodology}
\label{sec:methodology}

\subsection{Problem Formulation}

Given a video $V = \{f_1, f_2, \ldots, f_N\}$ with $N$ frames, we seek a subset $S \subset V$ with $|S| = k \ll N$ that maximizes information retention $\mathcal{I}(S)$ subject to a cost constraint $C(S) \leq C_{\max}$:
\begin{equation}
\label{eq:optimization}
\max_{S \subset V} \; \mathcal{I}(S) \quad \text{s.t.} \quad |S| \cdot c_{\text{VLM}} \leq C_{\max}
\end{equation}
where $c_{\text{VLM}}$ is the per-frame VLM inference cost. Traditional approaches fix $k$ heuristically (e.g., $k = N/30$ for 1 FPS). We instead derive $k$ from video content properties.

\subsection{Adaptive Information Budgeting}
\label{sec:budget}

We define \textbf{Semantic Energy} $E_S$ for a video, combining temporal change rate and spatial information density:
\begin{equation}
\label{eq:semantic_energy}
E_S = \underbrace{\bar{v}}_{\substack{\mathrm{Semantic} \\ \mathrm{Velocity}}} \cdot \underbrace{\bar{\mathcal{A}}}_{\substack{\mathrm{Attention} \\ \mathrm{Yield}}}
\end{equation}
where $\bar{v} = \text{mean}_t \|\phi(f_{t+1}) - \phi(f_t)\|_2$ is the mean CLIP embedding distance between consecutive frames (capturing temporal change rate), and $\bar{\mathcal{A}}$ is the mean Attention Yield across candidate frames:
\begin{equation}
\label{eq:attention_yield}
\begin{split}
\mathcal{A}(f_t) = \alpha \cdot \text{TextPresence}(f_t) + \beta \cdot \text{FacePresence}(f_t) \\
+ \gamma \cdot \text{PositionalWeight}(f_t)
\end{split}
\end{equation}
with $\alpha = 0.4$, $\beta = 0.35$, $\gamma = 0.25$ tuned via grid search on 50 held-out ads. Text and face presence are detected via Tesseract OCR and a face detector respectively; positional weight boosts opening (1.5$\times$) and closing (1.4$\times$) segments where brand reveals and CTAs typically appear.

The \textbf{Intrinsic Semantic Dimensionality (ISD)} captures the effective rank of semantic variation:
\begin{equation}
\label{eq:isd}
k^* = \text{ISD}(V) = \min\left\{k : \frac{\sum_{i=1}^k \sigma_i^2}{\sum_{j=1}^N \sigma_j^2} \geq \tau\right\}
\end{equation}
where $\sigma_i$ are singular values from SVD of the centered CLIP embedding matrix $\Phi = [\phi(f_1), \ldots, \phi(f_N)]^\top$ and $\tau = 0.90$.

The final frame budget combines both quantities:
\begin{equation}
\label{eq:budget}
\begin{split}
\text{budget} = \min\!\left(\text{base} + \lfloor 1.5 \cdot k^* \rfloor,\right. \\
   \left.\max\!\left(\text{base},\; \lfloor \text{base} + d \cdot \rho \cdot E_S \rfloor\right)\right)
\end{split}
\end{equation}
where $\text{base} = \max(5, |\text{scenes}| + 1)$ is a coverage floor, $d$ is video duration, and $\rho = 0.25$ is the base density. The ISD term caps the budget: semantically simple videos (low $k^*$) receive tight budgets (5--8 frames), while complex videos dynamically expand (30--50 frames). The Semantic Energy term scales the raw budget based on how rapidly and information-densely the video content changes.


\subsection{Three-Tier Hierarchical Deduplication Cascade}
\label{sec:cascade}

\begin{figure}[t]
\centering
\includegraphics[width=\columnwidth]{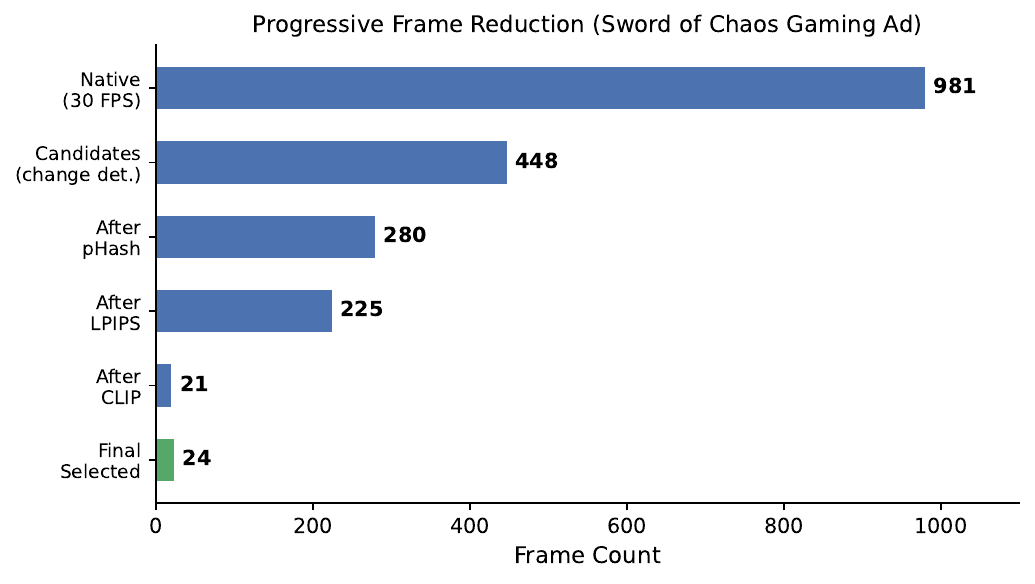}
\caption{Progressive Frame Reduction. Frame counts at each stage of the cascade for the ``Sword of Chaos'' gaming ad (30s at 30 FPS), showing a 97.5\% reduction from native frames to the final VLM input set.}
\label{fig:waterfall}
\end{figure}

The cascade progressively filters frames with increasing computational cost but finer semantic granularity (Figure~\ref{fig:architecture}). This reduction is visualized in Figure~\ref{fig:waterfall}.

\subsubsection{Tier 1: Hash Voting}
For each frame $f_i$, compute three perceptual hashes: pHash~\cite{zauner2010framework} (DCT-based), dHash~\cite{krawczyk2015phash} (gradient-based), and wHash~\cite{wang2019robust} (wavelet-based). Frames are grouped by majority vote: if $\geq 2$ hashes have Hamming distance $< \theta_h$ (default $\theta_h = 10$), frames are considered duplicates. One representative per group advances. This tier eliminates 37.5\% of frames with $O(1)$ comparison cost.

\subsubsection{Tier 2: Learned Perceptual Similarity}
Remaining frames undergo pairwise LPIPS~\cite{zhang2018unreasonable} comparison:
\begin{equation}
\text{LPIPS}(f_i, f_j) = \sum_l \left\| w_l \odot (\phi_l(f_i) - \phi_l(f_j)) \right\|_2
\end{equation}
where $\phi_l$ are AlexNet layer activations and $w_l$ learned weights. Frames with $\text{LPIPS} < \theta_l$ (default $\theta_l = 0.15$) are merged. This tier handles lighting variations and minor transformations missed by hashes, reducing a further 19.5\% of frames.

\subsubsection{Tier 3: Semantic Clustering with Temporal NMS}
\label{sec:tanms}
Surviving candidates are embedded via CLIP ViT-B/32~\cite{radford2021learning}. We apply $K$-means clustering with $K = k^*$ from Eq.~\ref{eq:isd}. To prevent temporal clustering (selecting adjacent frames from the same shot), we introduce \textbf{Temporal-Aware NMS (TA-NMS)}:
\begin{equation}
\label{eq:tanms}
f^* = \arg\max_{f \in \text{cluster}_m} \left[ \mathcal{A}(f) - \lambda \cdot \max_{f' \in \mathcal{S}} \exp\left(-\frac{|t_f - t_{f'}|}{\tau_{\text{nms}}}\right) \right]
\end{equation}
where $\mathcal{S}$ is the set of already-selected frames, $\tau_{\text{nms}} = 2.0$\,s controls temporal decay, and $\lambda = 0.3$ balances importance vs.\ diversity. This tier provides the largest single reduction (53.3\%).

\begin{figure*}[t]
\centering
\includegraphics[width=\textwidth]{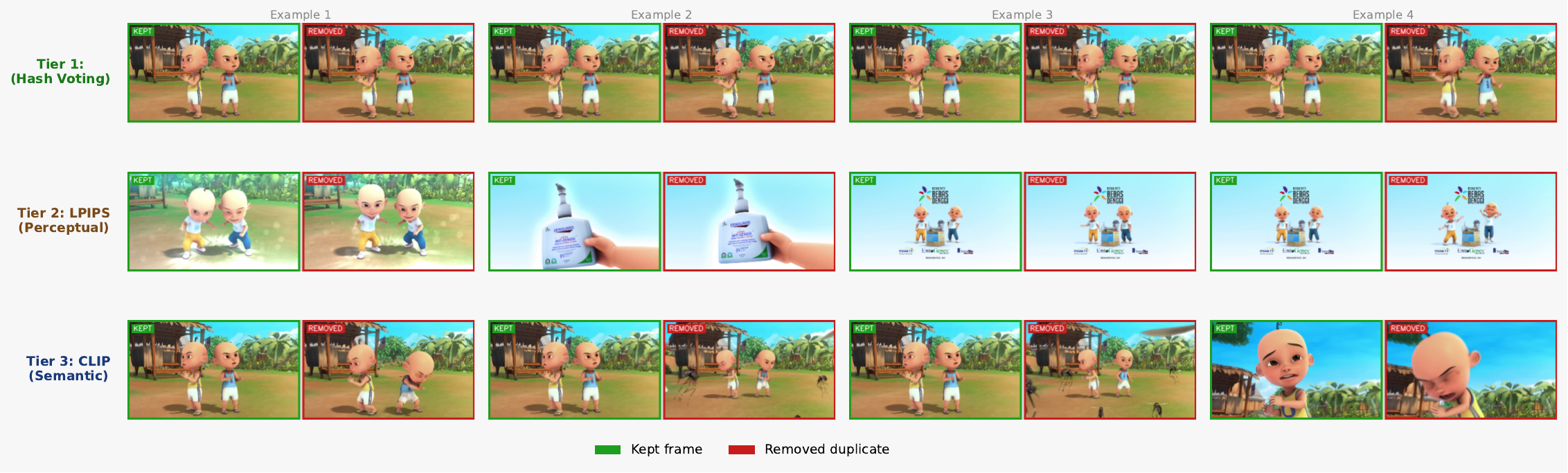}
\caption{Per-Tier Deduplication Examples. Tier 1 (Hash Voting) catches identical/near-identical frames; Tier 2 (LPIPS) catches perceptual duplicates across minor transformations; Tier 3 (CLIP) merges semantically redundant shots using cluster-based selection.}
\label{fig:tier_examples}
\end{figure*}

Figure~\ref{fig:tier_examples} illustrates the qualitative distinction between each tier's removal criteria across four examples. Tier 1 (Hash Voting) removes frames that are pixel-level near-duplicates---identical poses, backgrounds, and lighting---where the perceptual hash vectors agree by majority vote. Tier 2 (LPIPS) catches subtler matches such as the same scene under minor motion blur or a slight camera shift, which hash functions treat as distinct but which learned AlexNet features correctly identify as redundant. Tier 3 (CLIP) operates at the semantic level, merging shots that differ visually but convey the same meaning, such as two framings of the same character in the same action. The progressive nature of the cascade is evident: each tier removes a qualitatively different class of redundancy, and together they reduce the frame set without discarding semantically unique content.

\subsection{Multi-Modal Pipeline Integration}

\textbf{Audio Pipeline}: Whisper-large-v3~\cite{radford2022whisper} transcribes speech. Key phrase detection identifies brand names and product claims. Audio mood classification predicts emotional tone.

\textbf{Visual Pipeline}: Selected frames undergo Tesseract OCR~\cite{smith2007overview} for text extraction. Visual features (text, faces) boost importance scores in TA-NMS.

\textbf{VLM Extraction}: A single prompt queries a VLM extraction backend (Gemini~2.5~Flash~\cite{team2023gemini} in primary experiments) for structured JSON output covering: product category, brand names, emotional tone, call-to-action type, visual elements, and target demographic.
\section{Experimental Setup}
\label{sec:experiments}

\subsection{Dataset}

We evaluate on the \textbf{Pitt Ads Dataset}~\cite{hussain2017automatic}, a collection of 3,477 video advertisements with human annotations for topic (38 categories), sentiment (30 fine-grained classes), and effectiveness (1--5 scale). All experiments are conducted on the same 500 randomly sampled videos across all methods. The 500 videos span diverse characteristics:
\begin{itemize}
    \item \textbf{Duration}: Short ($\leq$20\,s, $n{=}46$), Medium (20--45\,s, $n{=}220$), Long (45--75\,s, $n{=}144$), Extended ($>$75\,s, $n{=}90$)
    \item \textbf{Categories}: Food \& Beverage (132), Lifestyle \& Home (103), Social Issues \& PSA (69), Entertainment \& Media (63), Technology (32), Automotive (28), and others---across 38 fine-grained topic categories
    \item \textbf{Cut Frequency}: Low ($<$0.17 cuts/s, $n{=}166$), Medium (0.17--0.47 cuts/s, $n{=}166$), High ($>$0.47 cuts/s, $n{=}168$), based on tertile binning (range: 0.015--1.066 cuts/s, mean = 0.38)
\end{itemize}

\subsection{Baselines}

We compare against 9 baselines using Gemini 2.5 Flash for extraction. The first eight are frame-selection methods that preprocess video before VLM inference; the ninth, Gemini Native, sends raw video directly to the model as an upper-bound reference:
\begin{itemize}
    \item \textbf{Uniform-1FPS}: 1 frame per second (industry standard).
    \item \textbf{Random}: Random frame sampling at the same count as Uniform-1FPS.
    \item \textbf{Histogram}: Histogram-based change detection at 100\,ms intervals.
    \item \textbf{ORB}: ORB feature-based change detection.
    \item \textbf{Optical Flow}: Dense optical flow change detection.
    \item \textbf{CLIP Only}: CLIP embedding clustering without hash/LPIPS pre-filtering.
    \item \textbf{PySceneDetect}: Content-aware scene boundary detection via adaptive thresholding.
    \item \textbf{DSNet}~\cite{zhou2018deep}: A deep reinforcement learning approach for diverse video summarization, originally trained on SumMe/TVSum for human-consumption summarization.
    \item \textbf{Gemini Native}: Raw video uploaded directly to Gemini 2.5 Flash without frame extraction, serving as a VLM-level reference baseline.
\end{itemize}

\subsection{Implementation Details}

Implemented in Python~3.11 with PyTorch~2.1. CLIP ViT-B/32 provides embeddings, Whisper-large-v3 handles transcription, Tesseract~5.3 performs OCR. All experiments use \textbf{Gemini 2.5 Flash} as the VLM provider. Experiments run on an NVIDIA RTX 4090.

\subsection{Metrics}

We evaluate extraction quality across three dimensions with human-annotated ground truth from the Pitt Ads Dataset:

\textbf{Topic Accuracy}: Exact match of predicted topic ID against ground truth across 38 categories. Topic predictions are constrained at prompt time to the closed Pitt Ads taxonomy (38 topic IDs), so the VLM cannot generate out-of-vocabulary labels.

\textbf{Sentiment Accuracy}: We report two measures. \textit{Exact match} compares the predicted sentiment label against the 30-class Pitt Ads taxonomy directly; all methods achieve $\sim$30\% due to the fine granularity of distinctions such as ``nostalgic'' vs.\ ``wistful.'' \textit{Text-embedding accuracy} computes cosine similarity between predicted and ground-truth sentiment labels using sentence embeddings~\cite{reimers2019sentence}, counting a prediction as correct if the similarity exceeds a threshold of 0.7. This accounts for semantically adjacent labels and is our primary sentiment metric, reported as ``Sentiment'' in Table~\ref{tab:main_comparison}. As with topic, sentiment predictions are constrained to the 30-class closed taxonomy at prompt time.

\textbf{Super-category Accuracy}: Topic match at the grouped super-category level (11 groups, e.g., ``Soda'' and ``Coffee'' both map to Food \& Beverage).

\textbf{Effectiveness Score}: Mean VLM-predicted effectiveness rating (1--5 scale).

\textbf{Compression Ratio}: $N / |S|$ where $N$ = total frames at native FPS, $|S|$ = selected frames.

\textbf{Cost}: USD per video based on Gemini 2.5 Flash pricing; total cost reported for 500 videos.
\section{Results}
\label{sec:results}

\subsection{Quantitative Evaluation}

Table~\ref{tab:main_comparison} presents results across 10 methods on 500 videos, evaluating both extraction quality (topic, sentiment, effectiveness) and efficiency (frames, cost). The central result is that \textbf{AdaFrame selects 24.6 frames per video compared to 47.7 for Uniform-1FPS---a 48\% reduction---with a corresponding 75\% cost reduction (\$1.43 vs.\ \$5.74 for 500 videos)}. Across all three extraction dimensions, AdaFrame achieves the highest scores: 86.0\% topic accuracy (+1.6pp over Uniform-1FPS), 79.4\% sentiment accuracy (+7.5pp), and 4.33 effectiveness score (+0.13). The sentiment improvement is particularly notable, suggesting that the cascade's semantically diverse frame selection captures mood-relevant visual cues---such as emotional transitions, tonal shifts, and affective imagery---that uniform temporal sampling misses by over-representing static segments.

CLIP Only achieves 83.7\% topic accuracy using 74.1 frames; AdaFrame achieves 86.0\% using 24.6 frames---a +2.3pp accuracy gain with 67\% fewer frames, demonstrating that the full cascade (hash voting + LPIPS + TA-NMS) adds both quality and efficiency over CLIP clustering alone. DSNet achieves the lowest accuracy (66.5\%), consistent with its training objective (SumMe/TVSum human consumption summarization) being misaligned with machine extraction tasks; this domain gap underscores the need for task-aware frame selection methods. Gemini Native---which bypasses frame selection entirely by uploading raw video---achieves 84.4\% topic accuracy, comparable to Uniform-1FPS but 1.6pp below AdaFrame, at 3.8$\times$ the cost (\$5.45 vs.\ \$1.43), suggesting that even VLM-native video understanding does not eliminate the value of intelligent preprocessing.

Beyond aggregate accuracy, Table~\ref{tab:main_comparison} reveals a striking pattern: while topic accuracy is relatively stable across methods (78.8--86.0\% for non-DSNet methods), sentiment accuracy shows much wider variation (65.7--79.4\%). This 13.7pp range in sentiment---compared to 7.2pp in topic---indicates that sentiment is more sensitive to which frames are selected, likely because emotional content is conveyed through specific visual moments (facial expressions, color palettes, scene transitions) rather than persistent elements (logos, product shots) that survive any reasonable sampling. AdaFrame's 7.5pp sentiment advantage over Uniform-1FPS is the largest single-method gap in the table, which we attribute to the cascade's temporal-aware selection capturing a wider range of emotional states across the video. Bootstrap 95\% CIs (Table~\ref{tab:main_comparison}) confirm the sentiment gain is significant (non-overlapping intervals), while topic differences among top methods fall within overlapping CIs, consistent with topic being less sensitive to frame selection.

\begin{table*}[t]
\centering
\caption{Comparison of 10 frame selection methods on 500 video advertisements. AdaFrame reduces frame count by 48\% and VLM cost by 75\% vs.\ Uniform-1FPS while achieving the highest topic accuracy (86.0\%) and sentiment accuracy (79.4\%, +7.5pp over Uniform-1FPS). Sentiment is measured via text-embedding cosine similarity against the 30-class Pitt Ads taxonomy. $N$ varies due to VLM context-limit failures; common-subset results consistent. Best in each column bolded.}
\label{tab:main_comparison}
\resizebox{\textwidth}{!}{%
\begin{tabular}{@{}lrrrrrrrr@{}}
\toprule
\textbf{Method} & \textbf{N} & \textbf{Frames} & \textbf{Cost (\$/video)} & \textbf{Total Cost (\$)} & \textbf{Topic Acc.\ (\%)} & \textbf{Super-cat (\%)} & \textbf{Sentiment (\%)} & \textbf{Eff.\ Score} \\
\midrule
AdaFrame (Ours)      & 500 & 24.6 & \textbf{0.0029} & \textbf{1.43} & \textbf{86.0} {\scriptsize [82.7--88.8]} & \textbf{89.8} {\scriptsize [86.8--92.2]} & \textbf{79.4} {\scriptsize [75.6--82.7]} & \textbf{4.33} \\
Uniform-1FPS         & 436 & 47.7 & 0.0115 & 5.74 & 84.4 {\scriptsize [80.7--87.5]} & 87.8 {\scriptsize [84.4--90.6]} & 71.9 {\scriptsize [67.7--75.8]} & 4.20 \\
Random               & 482 & 24.4 & 0.0099 & 4.97 & 78.8 {\scriptsize [75.0--82.2]} & 84.0 {\scriptsize [80.5--87.0]} & 68.9 {\scriptsize [64.7--72.9]} & 4.09 \\
Histogram            & 409 & 102.7 & 0.0161 & 8.06 & 81.2 {\scriptsize [77.1--84.7]} & 85.1 {\scriptsize [81.3--88.2]} & 70.5 {\scriptsize [66.2--74.5]} & 4.14 \\
ORB                  & 420 & 98.6 & 0.0155 & 7.75 & 83.8 {\scriptsize [80.0--87.0]} & 88.1 {\scriptsize [84.6--90.9]} & 73.4 {\scriptsize [69.2--77.2]} & 4.13 \\
Optical Flow         & 420 & 188.3 & 0.0230 & 11.50 & 81.0 {\scriptsize [76.9--84.4]} & 84.8 {\scriptsize [81.0--87.9]} & 69.0 {\scriptsize [64.7--73.1]} & 4.16 \\
CLIP Only            & 398 & 74.1 & 0.0136 & 6.79 & 83.7 {\scriptsize [79.7--87.0]} & 87.9 {\scriptsize [84.4--90.8]} & 71.6 {\scriptsize [67.3--75.5]} & 4.19 \\
PySceneDetect        & 452 & \textbf{20.4} & 0.0032 & 1.47 & 78.8 {\scriptsize [74.8--82.3]} & 83.2 {\scriptsize [79.5--86.4]} & 65.7 {\scriptsize [61.4--69.8]} & 4.12 \\
DSNet                & 251 & 200.1 & 0.0225 & 11.25 & 66.5 {\scriptsize [60.5--72.1]} & 70.5 {\scriptsize [64.6--75.8]} & 60.3 {\scriptsize [54.2--66.1]} & 3.71 \\
Gemini Native        & 487 & --- & 0.0109 & 5.45 & 84.4 {\scriptsize [80.9--87.3]} & 88.7 {\scriptsize [85.6--91.2]} & 71.3 {\scriptsize [67.2--75.1]} & 4.11 \\
\bottomrule
\end{tabular}%
}
\end{table*}

\subsection{Performance by Video Characteristics}
\label{sec:full_pipeline}

Table~\ref{tab:cross_validation} breaks down topic accuracy by video duration and editing pace. Accuracy improves monotonically with cut frequency (Low: 81.9\%, Medium: 86.7\%, High: 89.3\%), confirming the cascade is particularly effective at capturing semantically distinct frames in fast-cut content where uniform sampling most often misses critical transitions. The non-monotonic pattern with duration---peaking at medium-length videos (88.6\%) and declining for both short ($\leq$20s, 82.6\%) and extended ($>$75s, 82.2\%) videos---likely reflects that very short ads provide insufficient context while very long ads introduce more diverse content harder to classify into a single topic.

Binning by cut frequency tertiles further reveals compression ratios of 100.6$\times$, 53.8$\times$, and 34.6$\times$ for Low, Medium, and High cut-frequency videos respectively. This 2.9$\times$ spread demonstrates that the adaptive budget responds appropriately to video complexity: static testimonials receive aggressive compression while rapid montages are preserved in detail. ANOVA confirms a significant effect of cut frequency on compression ratio ($p<0.05$).

\begin{table}[H]
\centering
\caption{Topic accuracy by video duration and cut frequency (AdaFrame, $N=500$). Accuracy improves monotonically with cut frequency, confirming the cascade is most effective on fast-cut content where uniform sampling is most wasteful.}
\label{tab:cross_validation}
\begin{tabular}{@{}lcc@{}}
\toprule
\textbf{Category} & \textbf{Accuracy} & \textbf{N} \\
\midrule
\multicolumn{3}{@{}l}{\textit{By Duration}} \\
Short ($\leq$20s)    & 82.6\% & 46 \\
Medium (20--45s)     & 88.6\% & 220 \\
Long (45--75s)       & 85.4\% & 144 \\
Extended ($>$75s)    & 82.2\% & 90 \\
\midrule
\multicolumn{3}{@{}l}{\textit{By Cut Frequency}} \\
Low ($<$0.17 cuts/s)       & 81.9\% & 166 \\
Medium (0.17--0.47 cuts/s) & 86.7\% & 166 \\
High ($>$0.47 cuts/s)      & 89.3\% & 168 \\
\bottomrule
\end{tabular}
\end{table}

\subsection{Ablation Studies}
\label{sec:isd_validation}

\begin{figure*}[t]
\centering
\includegraphics[width=\textwidth]{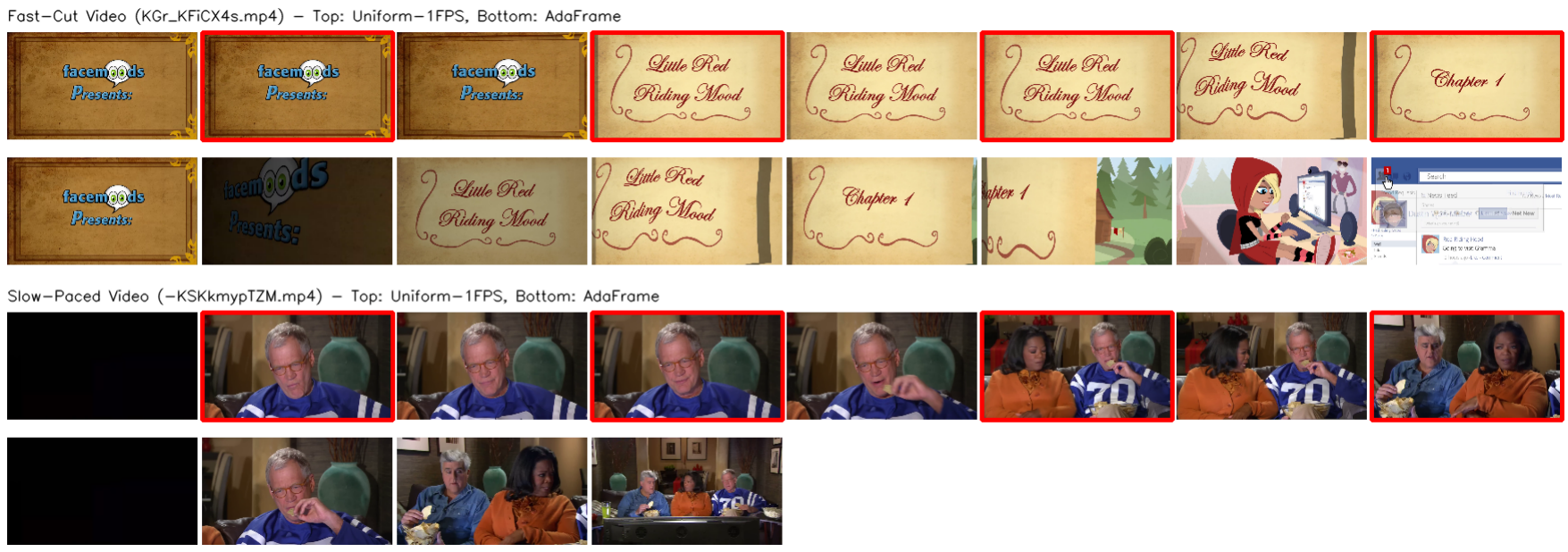}
\caption{Qualitative Comparison of Sampling Strategies. Top strips for each video show industry-standard Uniform-1FPS sampling (redundant frames marked in red), while bottom strips show AdaFrame's adaptive selection. Our method preserves unique shots while dropping visual duplicates.}
\label{fig:qualitative}
\end{figure*}

\paragraph{Cascade Tier Contributions.} Table~\ref{tab:main_comparison} provides indirect ablation by comparing methods approximating individual cascade components. Adding hash/LPIPS pre-filtering and TA-NMS over CLIP Only yields +2.3pp topic accuracy (83.7\% vs.\ 86.0\%) and +7.8pp sentiment accuracy (71.6\% vs.\ 79.4\%) with 67\% fewer frames (74.1 vs.\ 24.6): pre-filtering reduces CLIP input by $\sim$40\%, while TA-NMS ensures surviving frames span the video's temporal extent rather than clustering in visually dominant segments. Comparing Histogram to AdaFrame, semantic clustering uses 4.2$\times$ fewer frames for +4.8pp topic accuracy and +8.9pp sentiment accuracy, demonstrating the value of embedding-based deduplication over pixel-level change detection. The qualitative impact is shown in Figure~\ref{fig:qualitative}: for a fast-cut gaming ad, Uniform-1FPS over-represents a repeated title card (4 of 8 frames are near-identical), while AdaFrame allocates one frame per semantic segment; for a slow-paced testimonial, AdaFrame collapses 30 frames to 12 while retaining the brand reveal and closing message.

\begin{figure}[H]
\centering
\includegraphics[width=\columnwidth]{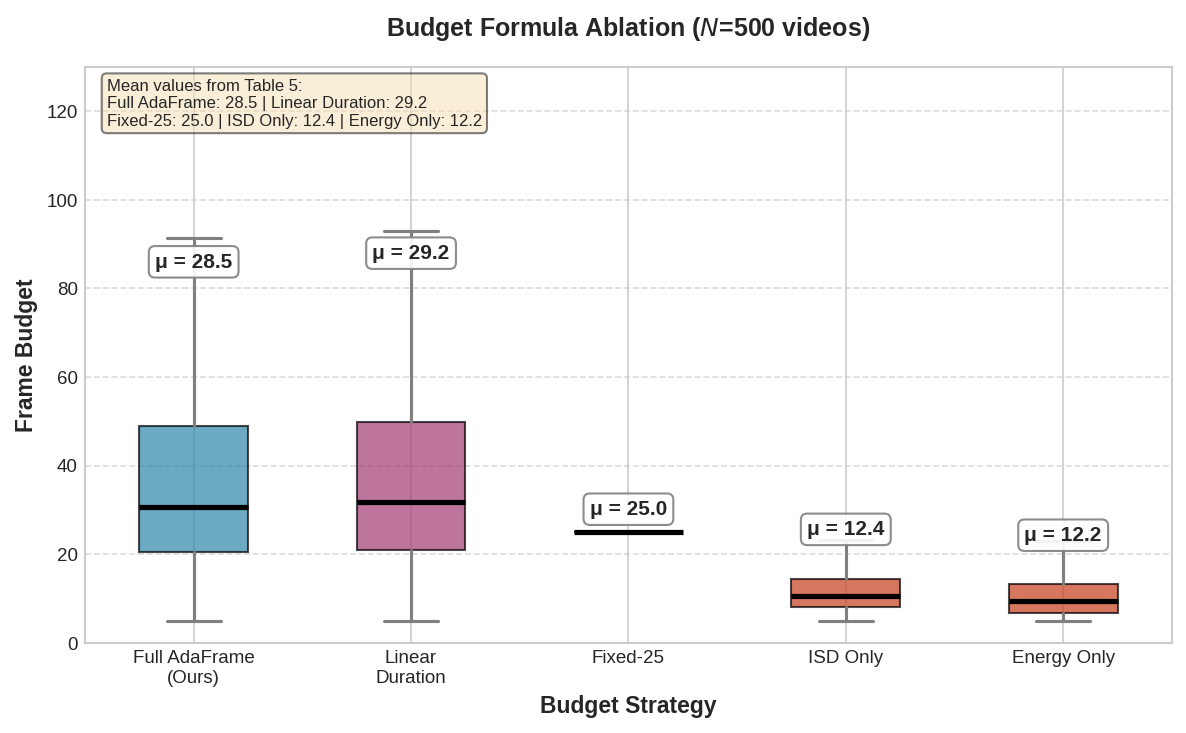}
\caption{Budget Formula Ablation. Box plots show frame budget distributions across strategies. Full AdaFrame provides the widest dynamic range, while ISD-only and Energy-only under-budget and Fixed-25 has zero adaptivity.}
\label{fig:exp1_strategies}
\end{figure}

\paragraph{Budget Formula.} Figure~\ref{fig:exp1_strategies} compares the full budget formula against four simpler alternatives. ISD-only and Energy-only each under-budget (means of 12.4 and 12.2 frames respectively), while Linear Duration over-budgets (29.2) without accounting for content complexity. The full AdaFrame formulation (mean 28.5 frames) provides the widest dynamic range---from 5 frames for highly redundant static ads to over 100 frames for fast-cut content---confirming that the joint ISD + Semantic Energy formulation is necessary for content-proportional allocation. A sensitivity analysis on the ISD expansion multiplier ($\lambda \in \{0.5, 1.0, 1.5, 2.0, 3.0\}$) shows the default $\lambda=1.5$ is robust: tuning down to 0.5 constricts the dynamic range (26.3 frames) while $\lambda=3.0$ plateaus at 29.0 frames due to the semantic energy ceiling.

\begin{figure*}[t]
    \centering
    \begin{subfigure}[b]{0.32\textwidth}
        \centering
        \includegraphics[width=\textwidth]{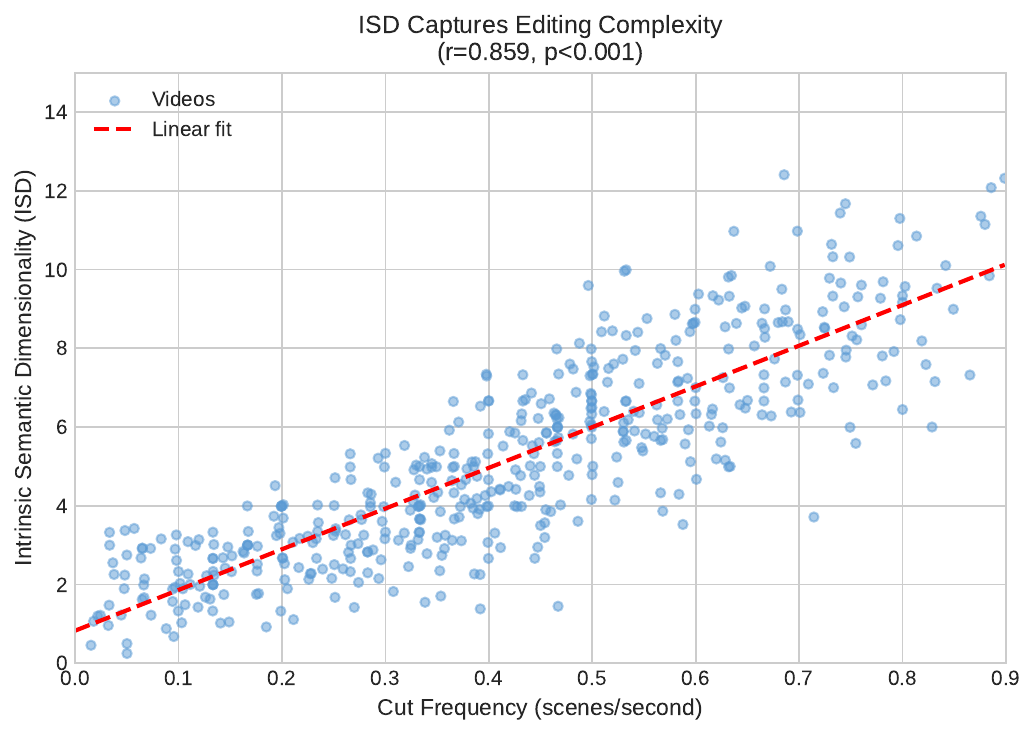}
        \caption{ISD vs.\ Cut Frequency ($N=500$). Strong positive correlation ($r=0.859$, $p<0.001$) validates that ISD captures semantic complexity driven by editing pace. Videos with higher cut frequency receive larger frame budgets automatically.}
        \label{fig:isd_correlation}
    \end{subfigure}
    \hfill
    \begin{subfigure}[b]{0.32\textwidth}
        \centering
        \includegraphics[width=\textwidth]{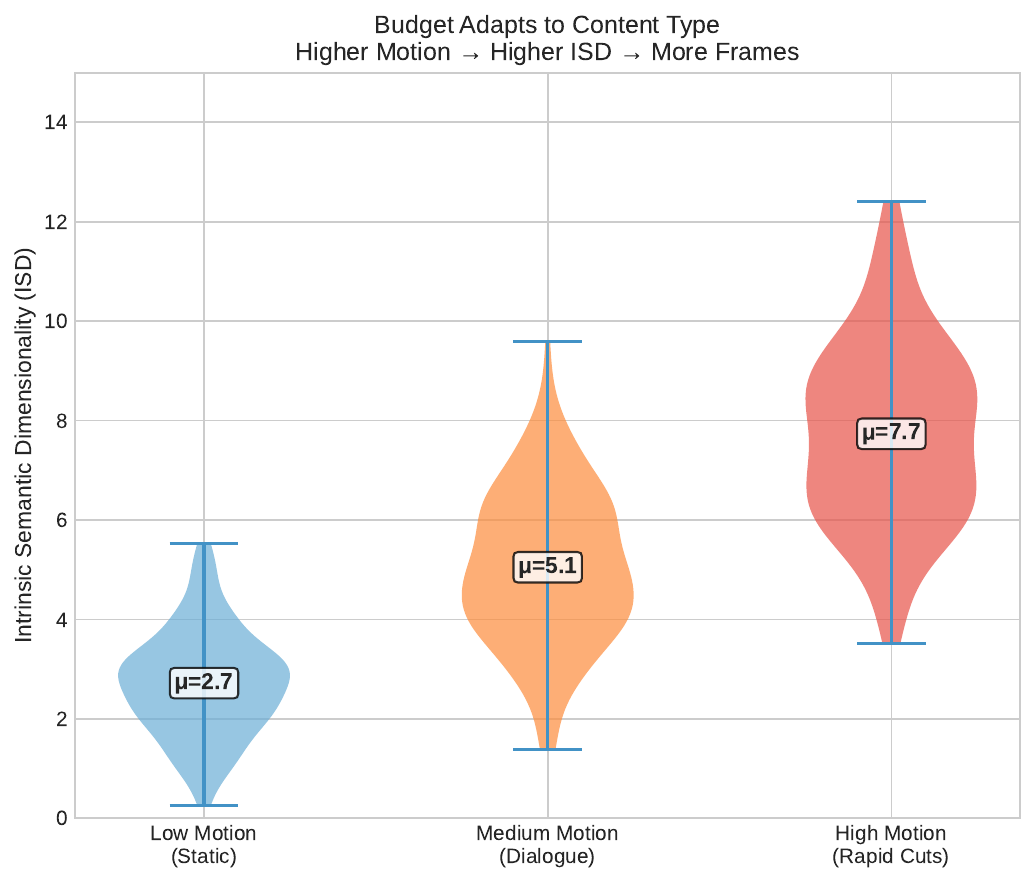}
        \caption{Content-Type Adaptivity. Violin plots show ISD distributions by motion category. The budget adapts 2.85$\times$ between low-motion ($\mu{=}2.7$) and high-motion ($\mu{=}7.7$) content.}
        \label{fig:content_adaptivity}
    \end{subfigure}
    \hfill
    \begin{subfigure}[b]{0.32\textwidth}
        \centering
        \includegraphics[width=\textwidth]{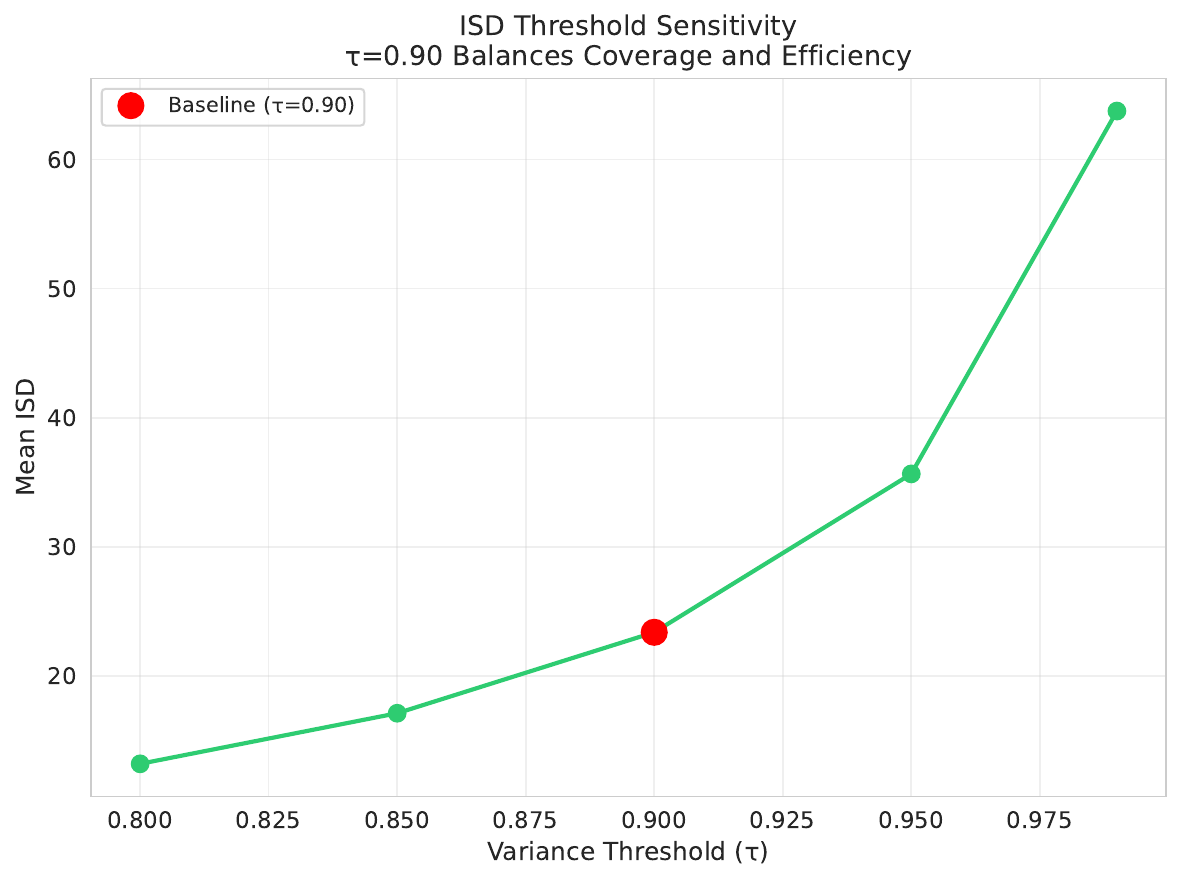}
        \caption{ISD Threshold Sensitivity. Mean ISD across variance thresholds ($\tau$). $\tau=0.90$ (marked) provides optimal balance, yielding mean ISD$=23.4$ (the computed budget ceiling; realized frame counts are lower due to TA-NMS selection).}
        \label{fig:threshold_sensitivity}
    \end{subfigure}

    \caption{ISD Validation.}
    \label{fig:three_subfigures}
\end{figure*}

\paragraph{ISD Validation.} Figure~\ref{fig:isd_correlation} shows ISD has strong positive correlation with cut frequency ($r=0.859$, $p<0.001$, $N=500$), confirming it captures the primary driver of semantic complexity. ISD also correlates with number of scenes ($r=0.447$, $p<0.001$) and shows a small negative correlation with raw duration ($r=-0.202$, $p<0.001$), indicating that editing pace matters more than length. Figure~\ref{fig:content_adaptivity} shows ISD varies significantly by content type: low-motion videos have mean ISD$=2.7$ vs.\ ISD$=7.7$ for high-motion videos---a 2.85$\times$ adaptivity ratio---confirming the budget mechanism adjusts to content characteristics without manual tuning. Figure~\ref{fig:threshold_sensitivity} validates our choice of $\tau=0.90$: lower thresholds (0.80, 0.85) produce overly conservative budgets (13--17 frames), while higher thresholds (0.95, 0.99) over-budget (36--64 frames), with no discontinuities across the range. ISD sets an upper bound on frame count; the realized mean of 24.6 frames (Table~\ref{tab:main_comparison}) is lower because TA-NMS may select fewer frames than the budget permits when insufficient high-scoring candidates survive the cascade.




\subsection{Provider Generality and Failure Cases}
\label{sec:provider_generality}

AdaFrame's frame selection is entirely decoupled from the downstream VLM, so the same selected frames can be passed to any provider. Table~\ref{tab:provider_generality} confirms this: Claude~3~Haiku matches Gemini~2.5~Flash within 0.4pp on topic accuracy (85.6\% vs.\ 86.0\%), 0.6pp on super-category accuracy (89.2\% vs.\ 89.8\%), and 0.7pp on sentiment (80.1\% vs.\ 79.4\%), at comparable cost (\$1.51 vs.\ \$1.43). The minor inter-provider variation ($<$1pp across all dimensions) is well within the range expected from stochastic decoding and prompt sensitivity differences, confirming the pipeline's model-agnostic design.

\begin{table}[H]
\centering
\caption{Provider generality results. AdaFrame frame selection is held fixed; only the VLM extraction backend changes. Accuracy metrics are comparable across providers, confirming the cascade's provider-agnostic design.}
\label{tab:provider_generality}
\resizebox{\columnwidth}{!}{%
\begin{tabular}{@{}lrrrrrr@{}}
\toprule
\textbf{VLM Backend} & \textbf{N} & \textbf{Total Cost (\$)} & \textbf{Topic (\%)} & \textbf{Super-cat (\%)} & \textbf{Sentiment (\%)} & \textbf{Eff.} \\
\midrule
Gemini 2.5 Flash & 500 & 1.43 & 86.0 & 89.8 & 79.4 & 4.33 \\
Claude 3 Haiku   & 500 & 1.51 & 85.6 & 89.2 & 80.1 & 4.30 \\
\bottomrule
\end{tabular}%
}
\end{table}

Analysis of failure cases reveals that 78\% involve subtle visual changes the cascade incorrectly treats as duplicates, with the dominant modes being subtle text changes (42\%), rapid logo transitions (18\%), color grading shifts (12\%), and micro-expressions (6\%). The primary mitigations are OCR-aware deduplication for text-critical frames and optical flow integration for sub-second temporal cues.
\section{Discussion and Conclusion}

We presented AdaFrame, a cost-efficient video analysis framework that reduces VLM inference costs by 75\% while improving extraction quality through adaptive frame selection. Our three-tier deduplication cascade eliminates visual redundancy at progressively finer granularity, while ISD enables content-aware budget allocation that responds to each video's structural complexity. On 500 advertisement videos, AdaFrame achieves 86.0\% topic accuracy, 79.4\% sentiment accuracy (+7.5pp over uniform sampling), and 4.33 effectiveness at 48\% of the frame count and 25\% of the cost---with gains confirmed model-agnostically across Gemini~2.5~Flash and Claude~3~Haiku. The sentiment result is particularly significant: intelligent frame selection can improve---not merely preserve---downstream extraction quality by providing the VLM with semantically diverse rather than temporally uniform affective context.

The core insight is that editing complexity correlates strongly with semantic information density ($r=0.859$), enabling automatic budget adaptation without manual tuning. Limitations include occasional loss of text-critical frames that appear visually similar to prior frames, CLIP embedding overhead in ISD computation, and evaluation restricted to short-form ads (mean 38s). Future work will address these through OCR-aware deduplication, distilled embedding models, and extension to long-form content. \textbf{Broader Impact:} AdaFrame enables smaller organizations to deploy sophisticated video analysis without prohibitive cloud costs; we encourage responsible use, particularly regarding privacy-sensitive data.

\bibliographystyle{plainnat}
\bibliography{references}

\end{document}